\documentstyle[12pt]{article}
\title{A note on Burgers' turbulence}
\author{S.\,Boldyrev\thanks{E-mail: boldyrev@princeton.edu}\\
{}~\\ 
{\em Department of Astrophysical Sciences, P.O. Box 451},\\
 {\em Princeton University, Princeton, NJ 08543}
}
\date{}
\begin{document}
\maketitle

\begin{abstract}
In this note 
the Polyakov
equation~[Phys.~Rev.~E~{\bf 52}\,(1995)\,6183] for the velocity-difference PDF,
with the exciting force correlation function $\kappa (y)\sim1-y^{\alpha}$ is
 analyzed. 
Several solvable cases are considered, which are in a good agreement 
with available
numerical results. Then it is shown how the method
developed by A.~Polyakov can be applied to turbulence with 
short-scale-correlated 
forces, a situation considered in models of self-organized criticality. 
\end{abstract}
\vskip1cm

\newpage

{\bf 1}. In Ref. [1] a method for calculating the velocity-difference PDF for 
Burgers' turbulence,
based on the conjectures on OPE, Galilean and scaling invariance, was proposed.
Starting with the Burgers' equation with a random Gaussian stirring force

\begin{eqnarray}
u_t + uu_x = \nu u_{xx} + f(x, t) \,\,,
\end{eqnarray}
$$
 \langle f(x,t)f( x^{\prime}, t^{\prime})
\rangle =\kappa (x-x^{\prime})\delta(t-t^{\prime})\,\,, \\
 \langle f(x,t) \rangle =0\,\,,
$$

~\\
\noindent it was obtained that the characteristic function for the 
velocity-difference PDF ($Z$-function), determined as

\begin{eqnarray}
Z(\mu, y) = \langle  \exp({\mu} [u(x+y/2)-u(x-y/2)]) \rangle \,\,,
\end{eqnarray}

\noindent obeys the following differential master equation:

\begin{eqnarray}
\left( \frac{\partial}{\partial \mu} - \frac{2b}{\mu} \right)\frac{\partial
Z}{\partial y} - (\kappa (0)-\kappa (y))\mu^2 Z = a(\mu) Z\,\,.
\end{eqnarray}

\noindent Here $\kappa(y)$ is the correlation function of the external force.  
The $\mu$~dependence of the $a$~anomaly must be chosen to conform 
to the scaling
invariance and can be different depending on the scaling properties of the force
correlation function. If for a large-scale-correlated force this function can be 
expanded as $\kappa (y) \sim 1-y^{\alpha}$, then  the $a$~anomaly must depend 
on $\mu$ as follows: 
$a(\mu)=a \mu^{\sigma} \,\,, \,\,\,\sigma = \frac{2-\alpha}{1+\alpha}$. 
Using the scaling ansatz $Z(\mu, y)=\Phi (\mu y^{\gamma})\,,\,\,\, 
\gamma = \frac{\alpha +1}{3}$, one can rewrite 
the equation~(3) in the form

\begin{eqnarray}
\gamma x\Phi^{\prime \prime} + \gamma (1-2b)\Phi^{\prime} - x^2 \Phi = 
a x^{\sigma}\Phi \,\,,
\end{eqnarray}

\noindent where $x=\mu y^{\gamma}$.

The unknown parameters $a$ and $b$ 
should be determined from
the main requirement that the PDF be a positive, finite and normalizable 
function. 
 Other possible restrictions for the theory are discussed later.

In this note we first
analyze  equation~(4) for arbitrary $\alpha$ in the force correlation 
function and
consider the following special solvable cases: $\alpha = 2$, 
$\alpha = 1$, $\alpha =
1/2$.  We then consider an application of the method of~[1] to  special kinds
of external forces -- forces correlated at short scales. The latter 
situation is
studied in various models of 
self-organized criticality and has wide applications~[2].

{\bf 2}. To begin with, 
we write down the asymptotics of the solutions of (4) for small $x$:

\begin{eqnarray}
\Phi(x) \sim 1 + \frac{a\gamma}{1-2b\gamma} x^{\frac{3}{\alpha + 1}} +
 c x^{2b} + ...\,\,,
\end{eqnarray} 

\noindent and for large positive $x$:

\begin{eqnarray}
\Phi(x) \propto \exp{\frac{2}{3\sqrt{\gamma}} x^{\frac{3}{2}}}\,\,,
\end{eqnarray}

\noindent where $a$, $b$, and $c$ should be
determined from the conditions mentioned above.

 We note that the most restrictive
condition, the condition of normalizability of the PDF, can be 
reformulated directly in terms 
of the
$\Phi$-function. Indeed, the function~$\Phi$ must be analytical 
in the right half of the complex 
plane~Re$x \geq 0$, and must
vanish for $x \rightarrow \rho \pm i\infty$, $\rho \geq 0$. This, along 
with the
condition of normalization $\Phi(0)=1$, gives the
quantization rule for $a$ and $b$.

Let us denote the Fourier transform of $\Phi(x)$ as ${\tilde w}(z)$, the 
velocity
distribution function being $w(u,y)={\tilde w}(u/y^{\gamma})/y^{\gamma}$. 
The integral
representation for ${\tilde w}(z)$ is:

\begin{eqnarray}
{\tilde w}(z) = \int \limits_{\rho -i\infty}^{\rho +i\infty} 
e^{-xz}\Phi(x)\,dx\,\,.
\end{eqnarray} 

The asymptotics of ${\tilde w}(z)$ for large positive $z$ is determined by 
 large $x$ and
is given by ${\tilde w}(z)\propto \exp(-\gamma z^3/3)$. To find the 
asymptotics 
for large
negative $z$, we deform the tails of the integration contour to coincide 
with the 
negative real axis. 
Since $e^{-xz}$  decays  rapidly as $x \rightarrow -\infty$, the
asymptotics is determined by the leading singularity in the expansion (5). 

In general,
two cases are possible. If $\frac{3}{\alpha + 1} < 2b$, the asymptotics is
$\tilde w \sim z^{-1 -\frac{3}{1+\alpha}}$. 
Such behavior is observed in numerical
simulations [3], 
which indicates that this inequality usually holds, and the $b$-anomaly
does not affect the asymptotics. For 
$2b < \frac{3}{\alpha + 1}$ the  asymptotics 
should in general be determined by the 
$b$~anomaly, ${\tilde w}(z)\sim z^{-2b-1}$, if~$2b$ is not an integer.

The asymptotics (5) also shows that there exist two degenerate cases. 
These are the cases when 
$\frac{3}{\alpha + 1}$ is an integer, and the corresponding term does not 
contribute to the
integral. These cases 
($\alpha =1/2,\,\,2$) are solvable and will be considered below.
We will also consider another solvable case, 
with $\alpha =1$.

{\bf 3}. We start with the case $\alpha = 2$. Let us Laplace-transform 
Eq.~(4) to get an equation for the probability distribution 
$w(u,y) = \tilde w(z)/y$~:

\begin{eqnarray}
\tilde w^{\prime \prime} + 
z^2 \tilde w^{\prime} +(1+2b)z \tilde w  = -a \tilde w \,\,,
\end{eqnarray}

\noindent where $z=u/y$, assuming the notation of~[1]. 
All derivatives in this equation are with respect to~$z$. 
Below we consider only the function
$\tilde w$ and drop the tilde sign. Asymptotics of the 
solution at $\vert z \vert \rightarrow \infty $ can be easily found from 
Eq.~(8):

\begin{eqnarray}
w \propto e^{-z^3/3} \,\,, \,\,\, w \sim \frac{1}{z^{2b+1}} \,\,\,.
\end{eqnarray}

\noindent We are looking for a 
physically reasonable solution, with the asymptotics

\begin{eqnarray}
& w \propto e^{-z^3/3} ,\,\,\, z \rightarrow +\infty\,\,, \nonumber \\
& w \sim \frac{1}{z^{2b+1}}, \,\,\, z \rightarrow -\infty \,\,.
\end{eqnarray}

For the $w$-function to be normalizable we should consider only $b > 0$. 
Upon writing $w = \Psi e^{-z^3/6}$, we exclude the first 
derivative from Eq.~(8) and get the Schr\"odinger equation for
the $\Psi$-function,

\begin{eqnarray}
-\Psi^{\prime \prime} + \left( \frac{z^4}{4} -2bz \right)\Psi = a \Psi \,\,,
\end{eqnarray}

\noindent mentioned in [1]. The ground state of this equation is a 
positive and
normalizable function. 
This is the only solution satisfying the general requirements
for the PDF. 
Thus, for any~$b>0$ we 
find the PDF as the ground state of the potential~(11), $a$~being
the energy of the ground state. 
Note that the case $b=1/2$ corresponds to the left tail of the
PDF~$\sim 1/u^2$, and the PDF 
obtained as a solution of (11), fits well the numerical
observations~[3] (see Figures). 
A numerical estimate in this case  gives for the $a$ anomaly  $a \simeq
0.354$.

An important remark should be made here. Integrating~Eq.~(8) from~$-\infty$ 
to~$+\infty$ for the case~$b>1/2$, we get: 
$$
(2b-1)\int z w(z)  = -a \int w(z) \,\,.
$$

We would like to stress that this expression does not contradict
 the requirement $\int w(u,y)u\,du = 0$. 
A significant contribution to the latter integral can
come from nonuniversal tails of the distribution function, not described by 
Eq.~(8). These tails are due to spontaneous breakdown of the Galilean 
symmetry~[1]. 
This fact should be taken into account when one 
compares the theoretical results with experimental observations.

Nevertheless a case exists 
for which $\langle z \rangle =0$, which corresponds to
$a=0$.  To consider it, we set~$a=0$ in~(8), and by the 
substitution~$s=-z^3/3$ 
 arrive 
at the degenerate hypergeometric equation:

\begin{eqnarray}
sw^{\prime \prime} + (\gamma  -s)w^{\prime} -
\alpha w=0\,\,,
\end{eqnarray}

\noindent with parameters $\gamma = \frac{2}{3}$, $\alpha =
\frac{1}{3}(2b+1)$.\footnote{Please, 
do not confuse these parameters, used only in
the analysis of (12), with the parameters $\alpha $ and
$\gamma $, introduced in (4).} 
The positive, finite and normalizable solution for this case has been 
found in~[1]. This solution can be constructed in the
following way: the only 
solution, exponentially decaying at $s \rightarrow -\infty$ 
and having power-like asymptotics at $s \rightarrow +\infty$, has 
the form

\begin{eqnarray}
& w(s) =\int \limits^{(s+)}_{-\infty} e^{t}\,(t-s)^{-\alpha}t^{\alpha -
\frac{2}{3}}\,dt\,\,,
 \,\,\, s < 0 \,\,,  \nonumber \\
& w(s) =\int \limits^{(0+)}_{-\infty} e^{t}\,(t-s)^{-\alpha}t^{\alpha -
\frac{2}{3}}\,dt\,\,, 
 \,\,\, s > 0 \,\,,
\end{eqnarray}

\noindent where in each integral the contour of the integration starts 
from $-\infty$, goes
around only one of the two singular points (denoted as the upper limits) 
in a positive direction and ends up at $-\infty$ again. 
One of these solutions can 
be analytically continued to the other one only if $\alpha
= n - 1/6$, where~$n$ is any integer number. It is interesting to note 
that this
exact quantization 
rule can be easily obtained as the Bohr-Sommerfeld condition 
for quantum
mechanics considered 
above, with zero energy.\footnote{This was pointed out by
V.~Gurarie} Positivity of the solution 
requires~$n=1$.

For the other degenerate case, the correlator of the
external force has the form: $\kappa (y) = 1 - y^{1/2}$. This force 
leads to a
differential equation for the $w(z)$-function, analogous to  Eq.~(8):

\begin{eqnarray}
w^{\prime \prime} + \frac{1}{2} z^2  
w^{\prime} +(\frac{1}{2}+b)z w  = a w^{\prime}\,\,,
\end{eqnarray} 

\noindent where $z=u/y^{1/2}$. Asymptotics of 
the left tail of the solution is given
by~(10). Excluding the first derivative from this equation, we
obtain the Schr\"odinger equation for the function 
$\Psi=w\exp (z^3/12 - az/2)$:

\begin{eqnarray}
-\Psi^{\prime \prime} + \left( \frac{z^4}{16} -\frac{a}{4}z^2 - 
bz \right)\Psi = 
-\frac{a^2}{4} \Psi \,\,.
\end{eqnarray}

As in the previous case, one can 
find the solutions as  the ground states of this equation. The
numerically observed PDF [3] has  
the left tail $\sim 1/u^3$ in the considered case. The same PDF
 can be obtained from our equation 
if we set $b=1$, i.e. when the $\beta$ anomaly is absent. One can
then numerically obtain $a \simeq -0.473$. 
A comparison of the whole PDF with the numerical results [3] shows 
a very good agreement (see Figures).

Equation  (15) has another interesting feature. For sufficiently 
large $a$ the 
potential is a two-well function, and one can show that the ground state is 
concentrated in the right well. 
Rescaling the variable~$z \rightarrow za^{-1/4}$ , one 
can expand the potential near the bottom of this well to obtain:

\begin{eqnarray}
-\Psi^{\prime \prime} + \frac{1}{2} z^2 \Psi = b {\sqrt 2} \Psi\,\,.
\end{eqnarray} 

\noindent It is interesting that $a$ has 
dropped out of this equation. Energy of the
ground state is $1/{\sqrt 2}$, which gives~$b=1/2$. 
A numerical estimate shows that
this result holds accurately already for~$a>1.5$. One could say that with
decreasing  energy, the system freezes at 
the point~$b=1/2$. For this value of~$b$
Eq.~(14) can be solved exactly.

To analyze the last case, $\alpha =1$, let us work in the 
$x$-representation.  
Note that by the substitution $\zeta =
x^{3/2}$ one can cast  equation (4) into the form:

\begin{eqnarray}
\frac{3}{2} \zeta \Phi^{\prime \prime} + \left( \frac{3}{2} 
- 2b \right)\Phi^{\prime}-
\left( \zeta + a \right)\Phi =0\,\,.
\end{eqnarray} 

This equation can be solved by the 
Laplace transform. The solution with the correct
asymptotics is:

\begin{eqnarray}
\Phi(x) = C x^{-\frac{3}{2}[\alpha_1 + \alpha_2 + 1]}
e^{\sqrt{\frac{2}{3}} x^{3/2}}
\int \limits_{-\infty}^{(0+)} e^{\tau}\, \tau^{\alpha_1}\left( \tau +
2\sqrt{\frac{2}{3}}x^{3/2} \right)^{\alpha_2}\, d\tau \,\,.
\end{eqnarray}

\noindent with

$$
\alpha_1 = - \frac{1}{2}\left[ 1+4b/3 \right] + a/\sqrt{6} \,,\,\,\,
\alpha_2 = - \frac{1}{2}\left[ 1+4b/3 \right] - a/\sqrt{6}\,\,.
$$

 $\Phi(x)$ will be an analytical function for Re$x\geq 0$, and a 
decaying function for
$x \rightarrow \rho \pm i\infty$ only when $\alpha_1 = n$ 
or $\alpha_2 = m$, where $n$
is any negative integer number and $m$ is any non-negative integer number. 
The only 
possibility of getting $\Phi(0)=1$ 
is $\alpha_1 = n$, which gives the following quantization rule: 
$a/\sqrt{6} -2b/3 = n +
1/2$. Positivity of the solution forces us 
to select $n=-1$, and (18) reduces to $\Phi =
\exp{\sqrt{\frac{2}{3}}x^{\frac{3}{2}}}$ .\footnote{V.~Gurarie 
informed me that this
solution can probably be obtained by the 
instanton method, along the lines of
[4].}

{\bf 4}. We now show how,  using the 
ideas of~[1], one can consider turbulence, excited by a
force correlated at small distances. We assume:

\begin{eqnarray}
 & \langle f(x,t)f( x^{\prime}, t^{\prime})\rangle = 
D(L,r_c)\delta(t - t^{\prime}) \,\,,\,\,  x= x^{\prime} \,\,, \nonumber \\
 & \langle f(x,t)f(x^{\prime}, t^{\prime} )\rangle = 0 
\,\,, \,\,\vert x- x^{\prime} \vert 
\gg r_c\,\,,
\end{eqnarray}

\noindent where $r_c$ is the correlation 
length, $L$ is the dimension of the system. 
We are interested in the velocity-difference PDF for large distances: $y \gg
r_c$. We shall assume that the integral of the force correlation function 
$\int \langle f(x,t)f( x^{\prime}, t^{\prime})\rangle dx$ remains 
finite as~$r_c \rightarrow 0$. In previous treatments of 
this problem, based mainly
on the renormalization group approach, the limit~$r_c \rightarrow 0$ was taken
before considering small~$\nu$ [2]. In fact, $\nu$ was kept finite there and 
entered the
final answer. Here we first take the limit $\nu \rightarrow 0$, and then 
consider the
second limit $r_c/y \rightarrow 0$. The answer 
should not depend on either~$\nu$
or~$r_c$. 

Assuming the Galilean and scaling 
invariance we obtain the following equation for 
the~$Z$~function:

\begin{eqnarray}
\left( \frac{\partial}{\partial \mu} - \frac{2b}{\mu} \right)\frac{\partial
Z}{\partial y} - D\mu^2 Z = a\mu^2 Z\,\,.
\end{eqnarray}

\noindent Now one can take 
the limit $r_c \rightarrow 0$. For this purpose consider the
singularity in the coefficient $D$:

\begin{eqnarray}
D(r_c, L) = 
\frac{1}{r_c} + O(1)\,,\,\,\,\, r_c \rightarrow 0\,\,.
\left( \frac{r_c}{L}\right)^n \,\,.
\end{eqnarray}

\noindent To get a finite limit, the singularity in the anomaly~$a$ should
cancel the same singularity in~$D$. Assuming such a cancellation we will
use the same letters~$D$ and~$a$ for the nonsingular parts of these
coefficients. Introducing a new function 
$\Phi$: $Z(\mu, y)=\Phi(\mu y^{1/3})$ we obtain an
equation for the Laplace transform of $\Phi$, completely 
analogous to the Eq. (8):

\begin{eqnarray}
3(D+a) w^{\prime \prime} + z^2  w^{\prime} +(1+2b)z  w  = 0  \,\,,
\end{eqnarray}

\noindent where $z=u/y^{1/3}$. Note that 
except for the cancellation of the singularity in the~D-coefficient, 
the~$a$-anomaly 
does not play an important role here. Introducing the new variable
$s=- \frac{z^3}{9(D+a)}$, we get exactly equation~(12). This equation has a
unique solution, 
which corresponds to the velocity-difference PDF with the left
tail~$1/u^{\frac{5}{2}}$.


Let us discuss 
the conditions for the Galilean and scaling invariance assumed in
the approach. The force~(19) generates finite $v_{rms}$:
 
$$v_{rms} = \lim_{r_c \rightarrow 0} \left[ r_c D(L, r_c)\right]^{1/3}\,\,,$$

\noindent  and the Galilean invariance holds for $u \ll v_{rms}$. 
The condition for the scaling invariance is $y\ll L$.

{\bf 5}. Finally, we discuss an 
important general restriction that can be imposed 
on the theory. This follows from the
physical condition of positivity of dissipation and was
proposed by A.~Polyakov.\footnote{A.~Polyakov, 1996, unpublished.} 
 It can be obtained if one notes that the
operator

\begin{eqnarray}
\frac{\partial^2}{\partial x^2} e^{(\lambda u(x) + \lambda_1 u(x_1) + ...)}
\end{eqnarray}

\noindent is not singular if $x$ does not coincide with any other $x_i$. 
Therefore, 

\begin{eqnarray}
\lim_{\nu \rightarrow 0} \nu \frac{\partial^2}{\partial x^2} e^{\lambda u(x)}
= 0 \,\,,
\end{eqnarray}

\noindent which leads to

\begin{eqnarray}
{ \alpha }(\lambda) Z + \frac{ \tilde \beta}{\lambda}  
\frac{\partial }{\partial x} Z = 
- \lim_{\nu
\rightarrow 0} \nu \langle \lambda^2 u_{x}^2 e^{\lambda u(x) + ...} \rangle  
\,\,, 
\end{eqnarray}

\noindent where we use the notation of [1]. The r.h.s. of this expression is 
negative. The function ${ \alpha} (\lambda)$ is analytical in the right 
half of the complex
plane, and may have a discontinuity at the
imaginary axis.  Summing up corresponding expressions 
for $\lambda_1 = \mu/2, \,\, x_1 = x+y/2$ and $\lambda_2 = -\mu/2, \,\, 
x_2 = x-y/2$,  
we get the following necessary 
condition, that must be valid for all non-negative~$x$:

\begin{eqnarray} 
a x^{\sigma} \Phi - 2(1-b)\Phi^{\prime} \leq 0 \,\,,
\end{eqnarray}

\noindent where $b= {\tilde \beta}+1$. 

One can easily see that this condition is  rather strong and allows one  to
considerably restrict the possible solutions of (3). For example, it prohibits 
the solutions 
with $b<3/4$ for the case $\alpha = 2$ and, probably, forces the $\beta$ 
anomaly 
to vanish for $\alpha
\leq 1/2$.

Nevertheless, one can see that this inequality is absent (or, at least, 
the above
arguments do not work) if we consider the dissipation in the form
$\nu \partial^{2p}/\partial x^{2p}$, with $p>1$ (the so-called 
hyperdissipation). 
This is the case 
for which the numerical simulations [3] have been performed.  One can expect 
that 
the structure of the shock front will be changed in this case. 
For example, one can show that for the dissipation of the form 
$\nu_1u_{xx}-\nu_2u_{xxxx}$ there exists the stationary solution, 
proportional to 
the linear combination of $-\tanh(x)$ and $-\tanh(x)/\cosh^2(x)$.

The intriguing question 
whether 
these different dissipations lead to different stationary regimes of the 
Burgers' turbulence 
and to what extent the found solutions can correspond to them has yet to 
be answered.

\vskip1cm

I am very grateful to A.~Polyakov for stimulating and interesting 
discussions and
suggestions. I would also like to thank V.~Gurarie for many useful discussions, 
V.~Yakhot for important conversations and for sharing with me the numerical 
results 
of [3], and J.~Krommes for drawing my attention to problems of self-organized 
criticality and for discussion of the results.

\vskip0.5cm
This work was supported by U.S.D.o.E. 
Contract No.~DE--AC02--76--CHO--3073.

\vskip2cm

\newpage
\underline{Figure captions.}\\
{}~\\
The following results of numerical simulations, 
 Figures 3 and 5, are taken from the paper by V.~Yakhot and 
A.~Chekhlov [3]. Fig. 3 shows the collapse of the PDFs in the universal 
region of $\Delta u$, and in our notation corresponds to $\alpha = 1/2$ in 
the force correlator. Fig. 5 shows the collapse of the PDFs in the universal 
region for $\alpha = 2$. For comparison, we have depicted the theoretical 
results by dots.

\end{document}